\documentclass[conference]{IEEEtran}
\usepackage{float}
\usepackage{cite}
\usepackage{amsmath,amssymb,amsfonts}
\usepackage{algorithmic}
\usepackage{graphicx}
\usepackage{textcomp}
\usepackage{xcolor}
\usepackage{tikz}
\usetikzlibrary{positioning,arrows.meta}
\def\BibTeX{{\rm B\kern-.05em{\sc i\kern-.025em b}\kern-.08em
    T\kern-.1667em\lower.7ex\hbox{E}\kern-.125emX}}
\begin{document}

\pagestyle{plain}  
\title{Grid Operational Benefit Analysis of Data Center Spatial Flexibility: Congestion Relief, Renewable Energy Curtailment Reduction, and Cost Saving}

\author{
\IEEEauthorblockN{
{\small
\begin{tabular}{ccc}
Haoxiang Wan & Linhan Fang & Xingpeng Li \\
\textit{Department of Electrical and} & 
\textit{Department of Electrical and} &
\textit{Department of Electrical and} \\
\textit{Computer Engineering} &
\textit{Computer Engineering} &
\textit{Computer Engineering} \\
\textit{University of Houston} &
\textit{University of Houston} &
\textit{University of Houston} \\
Houston, TX, USA & Houston, TX, USA & Houston, TX, USA \\
hwan6@cougarnet.uh.edu & lfang7@cougarnet.uh.edu & xli83@central.uh.edu
\end{tabular}
}
}
}

\maketitle

\thispagestyle{plain} 

\begin{abstract}
Data centers are computing infrastructures that process and store digital information. The rapid growth of artificial intelligence is driving a substantial increase in data center electricity demand, posing significant challenges to transmission networks. Geographically concentrated loads can trigger transmission congestion and reliability constraints, while the integration of renewable requires flexible resources to maintain grid reliability and minimize renewable curtailment. This paper proposes that data centers can act as grid resources by geographically reallocating computational workloads to address these challenges. An optimal power dispatch model demonstrates that spatially flexible data center demand can reduce operating costs and spinning reserve requirements, effectively enabling data centers to function as dispatchable load resources. Case studies on a modified IEEE 73-bus system show that inflexible data center placement can lead to severe transmission violations, with line overloads reaching up to 30.1\%. Enabling spatial flexibility mitigates these violations and restores system feasibility. Moreover, by strategically shifting demand toward solar-rich locations, spatial flexibility reduces solar curtailment by up to 61.0\%. These results indicate that spatially flexible data center demand can provide transmission-like operational value and enhance renewable energy utilization.
\end{abstract}


\vspace{-3mm}  
\section{Introduction}
\vspace{-1mm}  
\IEEEPARstart{T}{he} proliferation of artificial intelligence and big data analytics has catalyzed substantial growth in data center (DC) infrastructure \cite{b1}, with global facilities now accounting for approximately 3\% of total electricity supply \cite{b2}. This expansion coincides with accelerated renewable energy integration \cite{b3}, introducing operational complexities in transmission congestion management and economic dispatch optimization \cite{b4,b5,b6}. Recent advances in machine learning-based optimal power flow (OPF) formulations have enhanced computational efficiency while incorporating system-level constraints \cite{b7,b8}. According to Lawrence Berkeley National Laboratory's 2024 report, U.S. data centers consumed 4.4\% of electricity in 2023, with projections reaching 6.7--12\% by 2028 \cite{b9}. This AI-driven expansion introduces geographically concentrated loads that may intensify transmission stress and LMP volatility \cite{b10,b11}, requiring approaches beyond conventional transmission expansion with its substantial capital costs and extended timelines \cite{b12}.

Traditional congestion mitigation relies on physical infrastructure investments involving considerable capital and multi-year development cycles \cite{b13}. Demand-side management offers a complementary approach through load flexibility \cite{b14,b15}, yet existing frameworks predominantly address temporal flexibility with limited investigation of spatial flexibility---the strategic redistribution of computational workloads across geographically dispersed facilities. Data centers' distributed architectures and delay-tolerant workloads enable workload migration with minimal service impact, potentially functioning as spatially controllable demand resources \cite{b16}.

Recent literature has examined data center participation in ancillary services and demand response \cite{b14,b15}, generally treating facilities as localized flexible loads. However, their potential as network-aware assets influencing transmission congestion through spatial reallocation remains relatively unexplored. While Fridgen et al.~\cite{b16} demonstrated economic viability of spatial load migration for international market balancing, the transmission-level implications warrant further investigation.

While recent work by Lindberg et al.~\cite{b17} explored geographic load shifting driven by locational carbon signals, their approach primarily models data centers as autonomous entities optimizing against external signals. In contrast, this paper adopts a centralized co-optimization framework for analytical purposes, in which data center operators determine and offer limited operational headroom after internalizing their local operational and economic constraints. The grid-level operator then incorporates this pre-offered flexibility into a system-level optimization to ensure feasibility under network constraints. This perspective emphasizes the grid-reliability value of spatial flexibility, requiring explicit modeling of physical capacity constraints that is distinct from purely economic or carbon-driven load shifting. The concept of demand-side resources providing transmission-like services---where spatial load redistribution can substitute for or complement physical network capacity---has received limited systematic investigation. Existing studies have not fully characterized the techno-economic implications of data-center spatial flexibility for renewable curtailment mitigation, transmission congestion relief, and locational marginal price variability within integrated optimal power flow frameworks.

This paper presents an optimal power flow framework modeling spatially flexible data centers and evaluating their potential role in alleviating transmission congestion and renewable curtailment. Using a modified IEEE 73-bus test system with distributed data center capacity and high solar penetration, we examine four operational scenarios addressing: how inflexible deployment influences transmission feasibility, whether spatial flexibility restores operability without physical expansion, what transfer capability thresholds enable congestion mitigation, and to what extent demand-side flexibility reduces solar curtailment in transmission-constrained systems. The framework identifies critical feasibility thresholds, economic saturation points, and curtailment reduction potential. The findings suggest spatial reallocation offers operational benefits; however, practical implementation necessitates consideration of market design and coordination challenges beyond this study's scope.
\vspace{-2mm}

\section{Formulation of the Optimization Model}
\label{sec:formulation}
\vspace{-2mm}

This section details the integrated optimization model that co-optimizes conventional generation, security reserves, renewable resources, and flexible demand within an OPF framework.
\vspace{-2mm}
\subsection{Objective Function}
The objective minimizes total operational cost:
\begin{equation}
\label{eq:obj}
\min \sum_{g \in \mathcal{G}} c_g P_g ,
\end{equation}
where $P_g$ and $c_g$ denote the active power dispatch and marginal cost for generator $g \in \mathcal{G}$.

\vspace{-2mm}

\subsection{Network and Generation Constraints}

Conventional generation is bounded by capacity limits:

\begin{equation}
\label{eq:pg_limit}
P_g^{\min} \leq P_g \leq P_g^{\max}, \quad \forall g \in \mathcal{G},
\end{equation}

where $P_g$ is the real power output of generator $g$, $P_g^{\min}$ and $P_g^{\max}$ are minimum and maximum capacity limits, and $\mathcal{G}$ is the set of all generators.

Under Direct Current power flow approximation, active power flow $P_k$ on branch $k \in \mathcal{K}$ is:
\begin{equation}
\label{eq:flow_eq}
P_k = \frac{\theta_{f(k)} - \theta_{t(k)}}{x_k}, \quad \forall k \in \mathcal{K},
\end{equation}
where $\mathcal{K}$ is the set of transmission branches, $\theta_{f(k)}$ and $\theta_{t(k)}$ are voltage angles at the from-bus and to-bus terminals, $x_k$ is the series reactance, and $f(k), t(k)$ map branches to buses in set $\mathcal{N}$.

Branch flows are constrained by thermal ratings:
\begin{equation}
\label{eq:lineflow_limit}
-\text{RateA}_k \le P_k \le \text{RateA}_k, \quad \forall k \in \mathcal{K},
\end{equation}
where $\text{RateA}_k$ is the thermal capacity limit of branch $k$.

Nodal power balance at bus $n \in \mathcal{N}$ is:
\begin{multline}
\label{eq:balance}
\sum_{g \in \mathcal{G}(n)} P_g + P^{\text{solar}}_n
+ \sum_{k \in \mathcal{K}^-(n)} P_k
- \sum_{k \in \mathcal{K}^+(n)} P_k \\
= d_n + L_n^{\text{optimized}}, \quad \forall n \in \mathcal{N},
\end{multline}
·
where $\mathcal{G}(n)$ is the set of generators at bus $n$, $P^{\text{solar}}_n$ is dispatched solar generation, $\mathcal{K}^-(n)$ and $\mathcal{K}^+(n)$ are branches flowing into and out of bus $n$, $d_n$ represents the traditional inelastic electrical load, and $L_n^{\text{optimized}}$ denotes the optimized data center computational workload at bus $n$ after spatial redistribution.

\vspace{-2mm}
\subsection{Security and Reserve Constraints}
For $N\!-\!1$ reliability, let $r_g$ denote the spinning reserve provisioned by generator $g$, representing the additional upward active power capacity that can be delivered within the reserve response time window $\Delta T$. The spinning reserve $r_g$ for generator $g$ satisfies:
\begin{align}
P_g + r_g \leq P_g^{\max}, & \quad \forall g \in \mathcal{G}, \label{eq:reserve1}\\[3pt]
r_g \leq RU_g \cdot \Delta T, & \quad \forall g \in \mathcal{G}, \label{eq:reserve2}\\[3pt]
\sum_{m \in \mathcal{G}} r_m \geq P_g + r_g, & \quad \forall g \in \mathcal{G}, \label{eq:reserve3}
\end{align}
where $RU_g$ is the ramp-up rate parameter and $\Delta T$ is the reserve deployment window. Constraint \eqref{eq:reserve1} ensures dispatch plus reserve remain within capacity; \eqref{eq:reserve2} limits reserve by ramping capability; \eqref{eq:reserve3} guarantees system-wide reserve covers the loss of any unit.

\vspace{-1mm}
\subsection{Renewable and Flexible Demand Modeling}
The solar power output $P^{\text{solar}}_n$ at site $n$ is modeled with available maximum power $\overline{P}^{\text{solar}}_n$ and curtailment $P^{\text{curt}}_n$:
\begin{gather}
P^{\text{solar}}_n = \overline{P}^{\text{solar}}_n - P^{\text{curt}}_n, \label{eq:solar_dispatch} \\
0 \le P^{\text{curt}}_n \le \overline{P}^{\text{solar}}_n, \label{eq:curt_bounds}
\end{gather}
where $P^{\text{curt}}_n$ represents the curtailed solar power at site $n$. Equation \eqref{eq:solar_dispatch} explicitly models the inverter's dispatchability, allowing renewable output to be curtailed ($P^{\text{curt}}_n$) when transmission constraints prevent full absorption, providing a critical degree of freedom for feasibility analysis.

Data center workload flexibility is modeled through the following constraints:
\begin{gather}
L_n^{\text{optimized}} \ge (1-\beta)L_n^{\text{original}}, \quad \forall n \in \mathcal{N}_{\text{DC}}, \label{eq:transfer_ratio} \\[3pt]
L_n^{\text{optimized}} \le L_n^{\text{cap}}, \quad \forall n \in \mathcal{N}_{\text{DC}}, \label{eq:dc_capacity} \\[3pt]
\sum_{n \in \mathcal{N}_{\text{DC}}} L_n^{\text{optimized}} = \sum_{n \in \mathcal{N}_{\text{DC}}} L_n^{\text{original}}, \label{eq:load_conservation}
\end{gather}
where $L_n^{\text{original}}$ denotes the original data center load at bus $n$, 
$L_n^{\text{optimized}}$ represents the optimized data center load following spatial redistribution, 
$L_n^{\text{cap}}$ is the maximum data center capacity at bus $n$,
$\beta$ denotes the transferable workload ratio, 
and $\mathcal{N}_{\text{DC}} \subseteq \mathcal{N}$ represents the set of buses hosting data center loads. 

Constraint~\eqref{eq:transfer_ratio} defines the operational headroom of the facility. It ensures that a baseline portion $(1-\beta)$ of the workload—representing stateful, mission-critical tasks—remains local to satisfy Service Level Agreements (SLAs), while only the discretionary capacity (headroom) is exposed for spatial shifting. 
Constraint \eqref{eq:dc_capacity} enforces the physical power capacity limit of the facility infrastructure, and \eqref{eq:load_conservation} ensures system-wide conservation of total computational workload.

\vspace{-2mm}
\subsection{Model Discussion and Scalability Analysis}
This formulation \eqref{eq:obj}--\eqref{eq:load_conservation} represents a tractable linear programming framework that jointly optimizes energy generation, reserves, renewable curtailment, and flexible load dispatch. The model enables endogenous determination of economic dispatch with renewable spillage under network congestion and security considerations, while utilizing spatial flexibility of data center workloads to alleviate intermittency, moderate price fluctuations, and enhance overall system resilience.

The proposed formulation relies on specific modeling abstractions regarding data center flexibility. The transferable workload ratio, $\beta$, is interpreted not as total load fluidity, but as the operational headroom  within the facility.

Data center workloads vary in their migration characteristics. AI training workloads typically require checkpointing (saving model states) to enable migration. 
Checkpoint overheads depend on model scale, parallelization strategy, and storage pathways, and can be reduced to sub-second to a few-second time scales through localized and distributed checkpoint optimizations \cite{b18}. 
These time scales are substantially smaller than the minute-level rolling intervals used in real-time grid security dispatch. Moreover, transmission line thermal dynamics evolve over tens of minutes, with short-term emergency ratings commonly defined over 15--30 minute horizons. 
Accordingly, checkpoint--restart mechanisms operate on time scales that are conceptually compatible with power system security dispatch and thermal constraints.

Regarding inference, online interactive inference is latency-sensitive, whereas a substantial portion of batch workloads---including training, batch inference, and offline data processing are delay-tolerant and can be deferred or relocated without directly affecting user experience. For modeling purposes, we adopt a stylized workload composition in which these workload categories constitute heterogeneous portions of aggregate compute demand and differ in their degree of spatial flexibility due to checkpointing overhead, migration latency, bandwidth availability, and service-level constraints. The parameter $\beta$ is therefore introduced not as an empirical workload fraction, but as an abstract representation of net operational headroom, capturing the portion of aggregate workloads that operators may choose to spatially reallocate after internalizing migration costs and system constraints.

\vspace{-2mm}
\section{Case Studies}
\label{sec:casestudies}
\vspace{-1mm}  
This section evaluates spatial DC flexibility impacts on power system operations using a modified IEEE 73-bus test system augmented with three data centers with individual capacity limits of $L_n^{\text{cap}}=900$~MW. Each optimal power flow (OPF) problem is solved using Gurobi 11.0 under three representative single-period operating conditions, corresponding to peak, shoulder, and off-peak demand levels scaled to 120\%, 100\%, and 80\% of the base load, respectively. The four studies examine: (1) congestion impacts of inflexible DC deployment, (2) baseline feasibility under fixed DC placement, (3) flexibility threshold requirements, and (4) solar curtailment mitigation through spatial reallocation.
\vspace{-2mm}

\subsection{Case Study 1: Fixed DC Congestion Impact} 
\label{sec:case1}
\vspace{-1mm}
This study quantifies the congestion impacts of large-scale DC deployment by comparing the No-DC baseline against Fixed DC configurations. The Fixed DC scenario deploys 1,450~MW total capacity across three strategic locations: 870~MW at Bus~27, 290~MW at Bus~47, and 290~MW at Bus~73.

To diagnose infeasibility arising from concentrated DC loads, an $\alpha$-relaxation is applied to transmission line capacity constraints. Specifically, equation~\eqref{eq:lineflow_limit} is updated by:
\vspace{-2mm}
\begin{equation}
|P_k| \le \text{RateA}_k + \alpha_k, 
\quad \alpha_k \ge 0, 
\quad \forall k \in \mathcal{K}
\label{eq:alpha_relaxation}
\end{equation}

\vspace{-2mm}
where $\alpha_k$ denotes the virtual capacity (MW) added to branch $k$ to restore feasibility. By penalizing $\sum_{k} \alpha_k$ in the objective function, this formulation enables analysis under otherwise infeasible DC configurations and quantifies the magnitude of network constraint violations. Note that the $\alpha$-relaxation in \eqref{eq:alpha_relaxation} is applied exclusively to this infeasibility diagnosis case; subsequent cases with enabled flexibility enforce strict thermal limits ($\alpha=0$).

Table~\ref{tab:cs1_baseline} quantifies baseline system conditions and Fixed DC infeasibility. Under the No-DC baseline, Line~47 (Bus~27--33) operates at 33.9--40.6\% utilization across scenarios. When 1,450~MW of Fixed DC load is added, this line becomes critically overloaded, requiring 2.1--52.7~MW of virtual capacity to restore feasibility. The peak scenario demands 52.7~MW of relaxation—equivalent to 30.1\% above the line's thermal rating.
\vspace{-2mm}
\vspace{-4mm}
\begin{table}[H]
\renewcommand{\arraystretch}{1.1}
\caption{Baseline Line Flows (No-DC) and Infeasibility Diagnosis (Fixed DC)}
\label{tab:cs1_baseline}
\centering
\scriptsize
\setlength{\tabcolsep}{3.5pt}
\begin{tabular}{@{}l c | cc | cc@{}}
\hline
\textbf{Scenario} & \textbf{Line Cap.} & \multicolumn{2}{c|}{\textbf{No-DC Baseline}} & \multicolumn{2}{c}{\textbf{Fixed DC (Infeasible)}} \\
 & \textbf{(MW)} & \textbf{Flow (MW)} & \textbf{Util. (\%)} & \textbf{Status} & $\boldsymbol{\sum \alpha}$~\textbf{(MW)} \\
\hline
Peak      & 175 & 71.0 & 40.6 & Infeasible & 52.7 \\
Shoulder & 175 & 68.5 & 39.1 & Infeasible & 25.1 \\
Off-Peak & 175 & 59.3 & 33.9 & Infeasible & 2.1 \\
\hline
\end{tabular}
\vspace{-2mm}
\begin{flushleft}
\scriptsize
\textit{Note:} Line 47 (Bus 27$\rightarrow$33) is the critical bottleneck. $\alpha$: Virtual capacity required for feasibility restoration.
\end{flushleft}
\end{table}
\vspace{-3mm}
\vspace{-2mm}
To demonstrate the effectiveness of spatial DC flexibility, Table~\ref{tab:cs1_lineflow} compares Fixed DC violations against an Optimized DC configuration where flexibility parameter $\beta=0.5$ (i.e., 50\% of DC capacity can be spatially redistributed). Peak flows under Fixed DC reach 227.7~MW, exceeding line capacity by 30.1\%, while shoulder and off-peak scenarios exhibit 14.4\% and 1.2\% overloads, respectively. Spatial load redistribution with $\beta=0.5$ eliminates all violations, reducing peak utilization to 77.7\%, shoulder to 67.0\%, and off-peak to 39.7\%.

\vspace{-4mm}
\begin{table}[H]
\renewcommand{\arraystretch}{1.1}
\caption{Transmission Line Loading: Fixed vs. Optimized DC Deployment (Line 47: Bus 27$\rightarrow$33)}
\label{tab:cs1_lineflow}
\centering
\scriptsize
\setlength{\tabcolsep}{3.5pt}
\begin{tabular}{@{}l c | cc | cc@{}}
\hline
\textbf{Scenario} & \textbf{Cap.} & \multicolumn{2}{c|}{\textbf{Fixed DC}} & \multicolumn{2}{c}{\textbf{Optimized DC}} \\
 & \textbf{(MW)} & \multicolumn{2}{c|}{} & \multicolumn{2}{c}{$(\beta=0.5)$} \\
\cline{3-4} \cline{5-6}
 &  & \textbf{Flow (MW)} & \textbf{Util. (\%)} & \textbf{Flow (MW)} & \textbf{Util. (\%)} \\
\hline
Peak      & 175 & 227.7 & 130.1 & 136.0 & 77.7 \\
Shoulder & 175 & 200.1 & 114.4 & 117.2 & 67.0 \\
Off-Peak & 175 & 177.1 & 101.2 & 69.5  & 39.7 \\
\hline
\end{tabular}
\vspace{1mm}
\begin{flushleft}
\scriptsize
\textit{Notes:} (1) Util.: Utilization percentage relative to line capacity; (2) $\beta=0.5$ represents 50\% flexibility headroom available for spatial optimization.
\end{flushleft}
\end{table}
\vspace{-2mm}

Table~\ref{tab:cs1_system} summarizes system-wide economic impacts. Under the No-DC baseline, the system operates without data center loads, establishing reference operating costs and LMPs. The Fixed DC scenario results in infeasibility due to transmission congestion. The Optimized DC scenario with $\beta=0.5$ demonstrates that spatial load redistribution, while accommodating 1,450~MW of data center demand, increases operating costs but maintains feasibility. LMPs exhibit load-dependent variations, with shoulder and off-peak periods experiencing more pronounced increases (from \$19.64 to \$74.75 and \$15.89 to \$19.64, respectively) due to congestion-induced redispatch requirements.
\vspace{-3mm}
\vspace{-2mm}\vspace{-2mm}
\begin{table}[H]
\renewcommand{\arraystretch}{1.1}
\setlength{\tabcolsep}{2.5pt}
\caption{System-Wide Economic Impact: LMP and Operating Cost Analysis}
\label{tab:cs1_system}
\centering
\scriptsize
\begin{tabular}{@{}l | cc | c | cc@{}}
\hline
\textbf{Scenario} & \multicolumn{2}{c|}{\textbf{No-DC}} & \textbf{Fixed} & \multicolumn{2}{c}{\textbf{Optimized DC}} \\
 & \multicolumn{2}{c|}{} & \textbf{DC} & \multicolumn{2}{c}{$(\beta=0.5)$} \\
\cline{2-3} \cline{5-6}
 & \textbf{Cost} & \textbf{LMP} &  & \textbf{Cost} & \textbf{LMP} \\
 & \textbf{(\$)} & \textbf{(\$/MWh)} &  & \textbf{(\$)} & \textbf{(\$/MWh)} \\
\hline
Peak      & 119,288 & 74.75 & Infeas. & 228,058 & 75.64 \\
Shoulder & 62,521  & 19.64 & Infeas. & 131,808 & 74.75 \\
Off-Peak & 44,779  & 15.89 & Infeas. & 70,009  & 19.64 \\
\hline
\end{tabular}
\vspace{1mm}
\begin{flushleft}
\scriptsize
\textit{Notes:} (1) No-DC: Baseline without data center loads; Fixed DC: Spatially fixed deployment (infeasible); (2) Optimized DC ($\beta=0.5$): 50\% spatial flexibility.
\end{flushleft}
\end{table}
\vspace{-3mm}
\vspace{-2mm}
\vspace{-1mm}
The concentration of 870~MW DC capacity at Bus~27 renders Line~47 infeasible under all operating scenarios, with peak conditions requiring 52.7~MW of virtual capacity (30\% above line rating). Fixed DC deployment produces capacity overloads ranging from 1.2\% to 30.1\% across operating conditions. Spatial DC optimization with $\beta=0.5$ (representing 50\% redistribution flexibility), achieved by strategically redistributing computing loads across multiple network locations, eliminates all transmission violations while maintaining line utilizations between 39.7\% and 77.7\%. The findings indicate that geographically distributed load allocation could potentially contribute to alleviating transmission constraints and improving the integration of data centers.

\vspace{-2mm}

\subsection{Case Study 2: Baseline Flexibility Evaluation} 
\label{sec:case2}

\vspace{-2mm}

Fixed DC locations remain infeasible across all scenarios due to the 870~MW load concentration at Bus~27. In this case study, the minimum flexibility parameter is set to $\beta = 0.5$, allowing each DC to adjust its load by up to half of its initial load.
\vspace{-1mm}
The congestion relief achieved through spatial reallocation ranges from 2.1~MW (off-peak) to 52.7~MW (peak), as quantified by the $\sum \alpha$ metric in Table~\ref{tab:cs1_baseline}. Line~47 connecting Buses~27 and~33 consistently emerges as the critical transmission bottleneck across all operating conditions.
\vspace{-2mm}
\vspace{-1mm}
\vspace{-1mm}
\begin{figure}[H]
\centering
\includegraphics[width=0.62\linewidth]{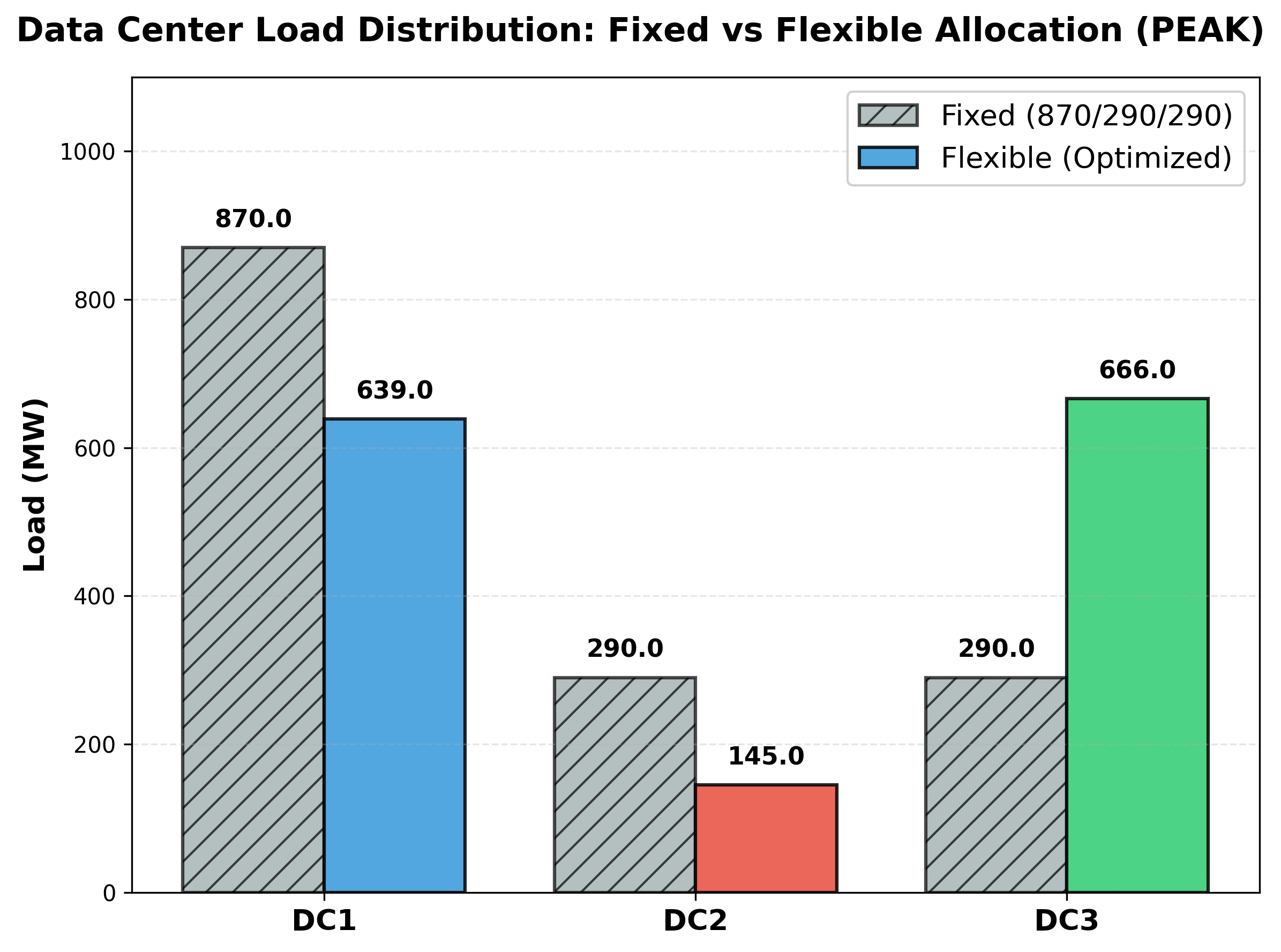}
\vspace{-1mm}
\caption{DC load distribution comparison: Fixed vs. Optimized deployment under flexibility constraints.}
\label{fig:dc_load_comparison}
\vspace{-1mm}\vspace{-2mm}
\end{figure}
\vspace{-3mm}
\vspace{-2mm}
Figure~\ref{fig:dc_load_comparison} compares the data center load distributions under fixed and optimized deployment across three operational scenarios. During peak conditions, spatial optimization redistributes loads such that DC$_1$ reduces from 870~MW to 639~MW, DC$_2$ decreases to 145~MW, and DC$_3$ increases to 666~MW. In the shoulder period (base case), DC$_1$ further decreases to 618.7~MW, DC$_2$ maintains 145~MW, and DC$_3$ reaches 686.3~MW. The off-peak scenario exhibits distinct reallocation patterns: DC$_1$ decreases to 565.6~MW, while DC$_2$ experiences substantial increase to 657.9~MW and DC$_3$ reduces to 226.5~MW. All scenarios preserve the total DC load of 1,450~MW as enforced by constraint~\eqref{eq:load_conservation}.
\vspace{-2mm}
\vspace{-1mm}
\subsection{Case Study 3: Flexibility Threshold Analysis}
\label{sec:case3}
\vspace{-1mm}
This study analyzes the impact of DC transferable workload ratio ($\beta$) on system economics, varying from $\beta=0.1$ to $\beta=0.5$. The system includes three DCs totaling 1,450~MW (870~MW at Bus~27; 290~MW each at Buses~47 and 73). Five discrete $\beta$ values are examined under peak, shoulder, and off-peak loads.

\begin{figure}[H]
\centering
\includegraphics[width=0.6\columnwidth]{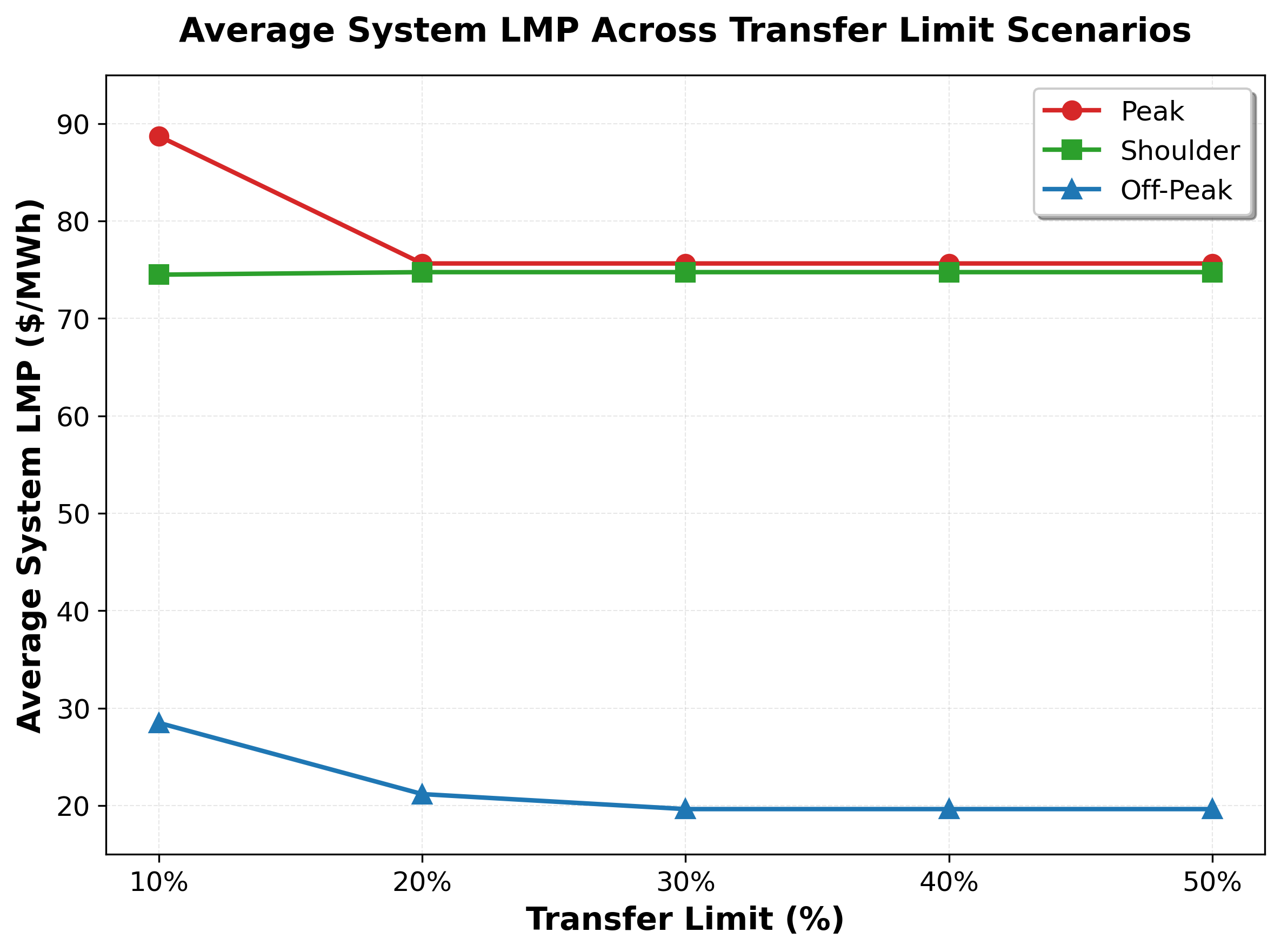}
\caption{Average System LMP Across Transfer Limit Scenarios}
\label{fig:cs3_lmp}
\end{figure}

Table~\ref{tab:cs3_cost} shows diminishing marginal cost savings as $\beta$ increases. Peak and Shoulder scenarios reach saturation at $\beta=0.2$, saving \$3,794 and \$135 respectively. Off-Peak requires $\beta=0.3$ for saturation with \$4,913 savings. Beyond these thresholds, further increases yield negligible benefits.
\vspace{-1mm}
\vspace{-3mm}
\begin{table}[H]
\renewcommand{\arraystretch}{1.1}
\caption{Total System Cost Under Varying DC Transfer Limits}
\label{tab:cs3_cost}
\centering
\scriptsize
\begin{tabular}{@{}lccc@{}}
\hline
Transfer Limit ($\beta$) & Peak (\$) & Shoulder (\$) & Off-Peak (\$) \\
\hline
$\beta=0.1$ & 231,852.40 & 131,942.55 & 74,922.17 \\
$\beta=0.2$ & 228,057.73 & 131,808.16 & 70,094.19 \\
$\beta=0.3$ & 228,057.73 & 131,808.15 & 70,008.95 \\
$\beta=0.4$ & 228,057.73 & 131,808.15 & 70,008.95 \\
$\beta=0.5$ & 228,057.73 & 131,808.16 & 70,008.95 \\
\hline
\end{tabular}
\end{table}
\vspace{-2mm}
\vspace{-1mm}
Figure~\ref{fig:cs3_lmp} illustrates system-wide LMP response. While higher $\beta$ generally reduces LMPs, the Shoulder scenario (green squares) shows distinct behavior. Average LMP remains constant at \$74.75/MWh as $\beta$ increases from 0.1 to 0.2, despite the \$135 cost reduction. This reflects that marginal cost (LMP) is determined by the marginal generating unit, while total cost depends on dispatch volumes of all infra-marginal units. Load redistribution from Bus~27 reduces expensive unit dispatch without changing which unit sets the marginal price. Peak and Off-Peak scenarios show significant LMP reductions, indicating flexibility displaced binding constraints.

Peak and Shoulder stabilize at $\beta=0.2$, while Off-Peak requires $\beta=0.3$ (curve flattening in Fig.~\ref{fig:cs3_lmp}). This finding—lower load requiring higher flexibility—occurs because reduced loading enables more aggressive redistribution before hitting alternative constraints. The non-monotonic relationship demonstrates topology dependence: specific DC placement creates congestion patterns varying with loading conditions. Quantitative conclusions (e.g., 20-30\% flexibility threshold) are contingent on system topology and DC locations, indicating deployment strategies must consider diverse operating scenarios.

\vspace{-2mm}
\vspace{-2mm}\vspace{-1mm}\vspace{-1mm}
\vspace{-1mm}
\vspace{-1mm}
\subsection{Case Study 4: Solar Curtailment Management}
\label{sec:case4}
\vspace{-1mm}
The modified IEEE 73-bus system includes 2{,}800~MW of solar PV located near congested corridors (buses~27 and~33) and three data centers totaling 1{,}450~MW of demand, each with up to 50\% spatial transferability ($\beta=0.5$). Two operating cases are considered: Fixed DC with inflexible placement, and Optimized DC allowing spatial reallocation toward solar-rich buses.

\vspace{-1mm}
\begin{figure}[H]
\centering
\includegraphics[width=0.7\linewidth]{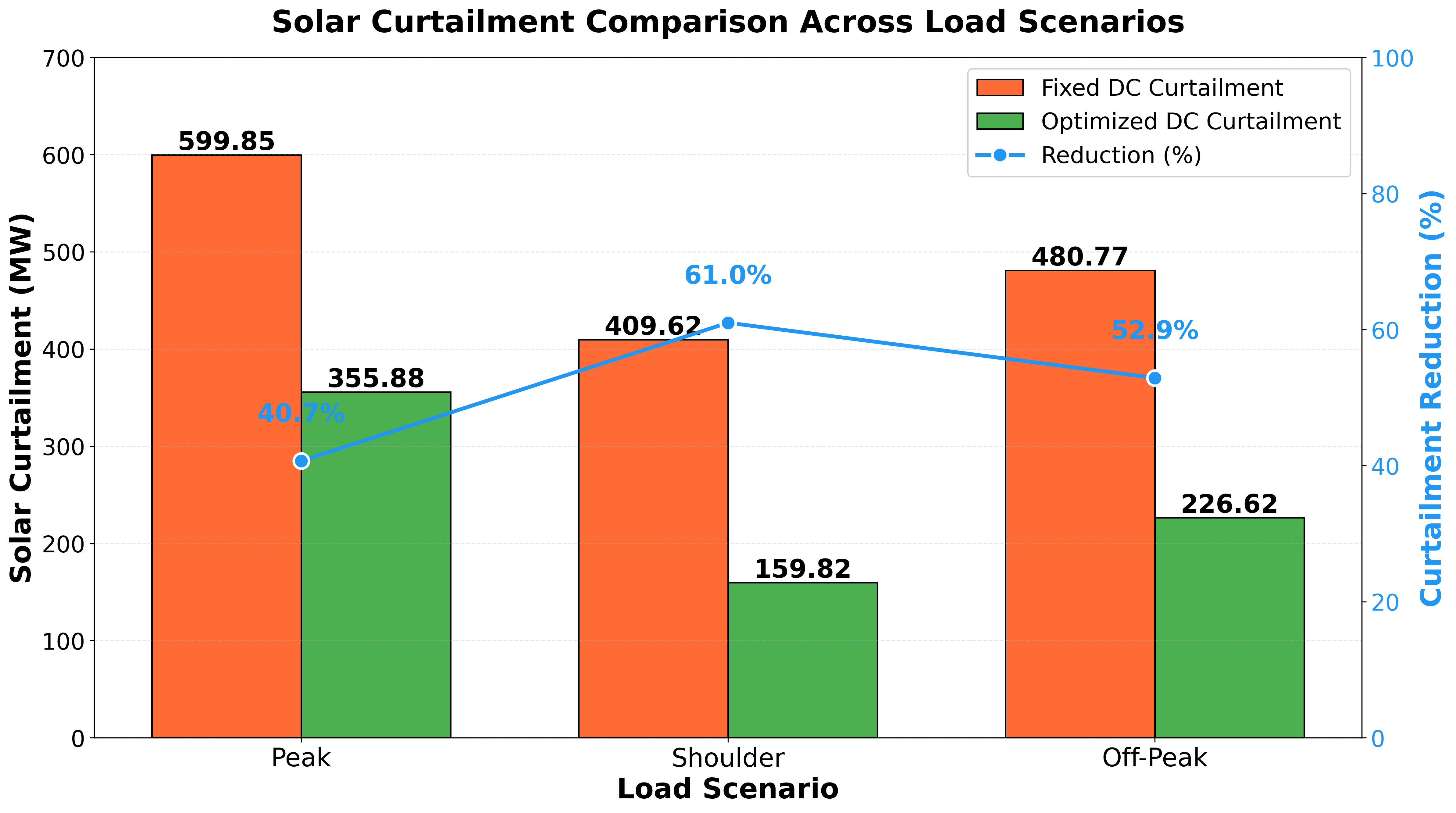}
\vspace{-3mm}
\caption{Solar curtailment comparison across load scenarios under fixed and optimized DC operation.}
\label{fig:solar_curtailment}
\vspace{-1mm}
\end{figure}
\vspace{-1mm}\vspace{-1mm}
Figure~\ref{fig:solar_curtailment} summarizes solar curtailment under peak, shoulder, and off-peak conditions. With Fixed DC deployment, curtailment reaches 599.85~MW, 409.62~MW, and 480.77~MW, primarily due to transmission congestion near buses~27 and~33. Enabling spatially optimized DC operation reduces curtailment to 355.88~MW, 159.82~MW, and 226.62~MW, corresponding to reductions of 40.7\%, 61.0\%, and 52.9\%, respectively, indicating improved utilization of locally available solar resources under transmission constraints.

\vspace{-1mm}
\vspace{-1mm}
\section{Conclusion}
\label{sec:conclusion}
\vspace{-1mm}
An OPF-based assessment on a modified IEEE 73-bus system shows that inflexible, concentrated data-center deployment can cause severe transmission congestion and infeasibility, with overloads up to 30.1\%. Introducing spatial workload flexibility restores feasibility and reduces solar curtailment by up to 61.0\% through load reallocation toward resource-rich locations, while 20--30\% transferability captures most of the operational cost benefits.

While this study establishes a deterministic benchmark for the technical potential of spatial flexibility, practical deployment requires bridging the gap between theoretical optima and grid stochasticity. Future work must extend this framework to incorporate the stochastic nature of renewable forecasts, integrate unit commitment logic, and analyze coupled transmission--distribution coordination \cite{b19}. Such extensions are essential for developing practical, co-optimized flexibility resources to support deep decarbonization pathways.

\end{document}